\documentclass[prl,
twocolumn,
showpacs,
tightenlines,
nofootinbib]{revtex4}
\usepackage{graphicx}
\usepackage{epsf}
\usepackage{amsmath}
\usepackage{mathrsfs}
\usepackage{xcolor}
\usepackage{cancel}
\usepackage{nicefrac}
\def\beq{\begin{equation}}
\def\eeq{\end{equation}}
\def\bea{\begin{eqnarray}}
\def\eea{\end{eqnarray}}
\def\beqa{\begin{equation}\begin{array}{l}}
\def\eeqa{\end{array}\end{equation}}
\def\eqlab#1{\label{eq:#1}}
\def\figlab#1{\label{fig:#1}}

\def\eref#1{(\ref{eq:#1})}
\def\Eqref#1{Eq.~(\ref{eq:#1})}

\def\Figref#1{Fig.~\ref{fig:#1}}

\def\half{\mbox{\small{$\frac{1}{2}$}}}

\def\quarter{\mbox{\small{$\frac{1}{4}$}}}

\def\barr{\left(\begin{array}{c}}
\def\earr{\end{array}\right)}
\def\bmat{\left(\begin{array}{cc}}
\def\emat{\end{array}\right)}
\def\al{\alpha}
\def\be{\beta}
\def\ga{\gamma} 
 
\def\veps{\varepsilon}  \def\eps{\epsilon}

\def\si{\sigma} \def\Si{{\it\Sigma}}
\def\th{\theta}  
\def\w{\omega}

\def\pa{\partial}

\def\pa{\partial}

\def\nn{\nonumber}

\def\lag{{\mathscr L}}

\def\cE{\mathscr{E}}

\def\bE{{\bf E}}
\def\bB{{\bf B}}

\def\3d{3-D}

\def\ol#1{\overline{#1}}

\begin{document}

\preprint{MITP/13-023}
\title{Separation of proton polarizabilities with the beam asymmetry of Compton scattering }

\author{Nadiia Krupina}
%\affiliation{Institut f\"ur Kernphysik, Johannes-Gutenberg Universit\"at, Mainz, Mainz D-55099, Germany}

\author{Vladimir Pascalutsa}

\affiliation{PRISMA Cluster of Excellence   Institut f\"ur Kernphysik, Johannes Gutenberg--Universit\"at Mainz, 55128 Mainz, Germany}

\date{\today}

\begin{abstract}
We propose to determine the magnetic dipole polarizability
of the proton from the
beam asymmetry of low-energy Compton scattering
based on the fact that the leading non-Born contribution to the asymmetry
is given by the magnetic polarizability alone;
the electric polarizability cancels out. The beam asymmetry thus provides
a simple and clean separation of the magnetic
polarizability from the electric one. Introducing
polarizabilities in a Lorentz-invariant fashion, we 
compute the higher-order (recoil) effects of polarizabilities
on beam asymmetry and show that these effects are
suppressed in forward kinematics. With
 the prospects of precision Compton experiments 
at the MAMI and HIGS facilities in mind, we argue why
the beam asymmetry could be the best way to measure
the elusive magnetic polarizability of the proton.
\end{abstract}


\maketitle
\thispagestyle{empty}
%\tableofcontents

%\section{Introduction}
The current Particle Data Group (PDG)~\cite{Beringer:1900zz}
values of the electric- and
magnetic-dipole polarizabilities of the proton~\cite{Baldin,Holstein:1992xr},
i.e.,
\begin{subequations}
\bea
\alpha_{E1} &=& (12.0\pm 0.6)\times 10^{-4}\,\mbox{fm$^3$}, \\
\beta_{M1} &=& (1.9\pm 0.5)\times 10^{-4}\,\mbox{fm$^3$}
\eqlab{PDGbeta}
\eea
\end{subequations}
are in significant disagreement with the most recent postdictions
of chiral effective field theory (ChEFT) 
\cite{Lensky:2009uv,McGovern:2012ew}, as can be seen in Fig 1.
The state-of-the-art ChEFT calculations, based
on either the baryon (B) or heavy-baryon (HB) chiral
perturbation theory (ChPT) with octet 
and decuplet fields~\cite{Lensky:2012ag},
are in excellent agreement with the experimental
Compton-scattering cross sections, but not
necessarily in agreement with the polarizabilities
extracted from these data by the experimental groups,
c.f.~\cite{Griesshammer:2012we} for review.
The situation is becoming more acute as the demand for precise
knowledge of nucleon polarizabilities is growing; they 
are for instance the main source of uncertainty in the extraction
of the proton charge radius from the muonic 
hydrogen Lamb shift (see  \cite{Pohl:2013yb} for a recent review).

%\parbox{.4\linewidth}{%
\begin{figure}[b]
%\begin{minipage}[c]{.4\linewidth}
%\vskip1mm
%\centerline{\epsfclipon  \epsfxsize=6cm%
  %\epsffile{olfig62.eps} 
%}
\includegraphics[width=0.8\linewidth]{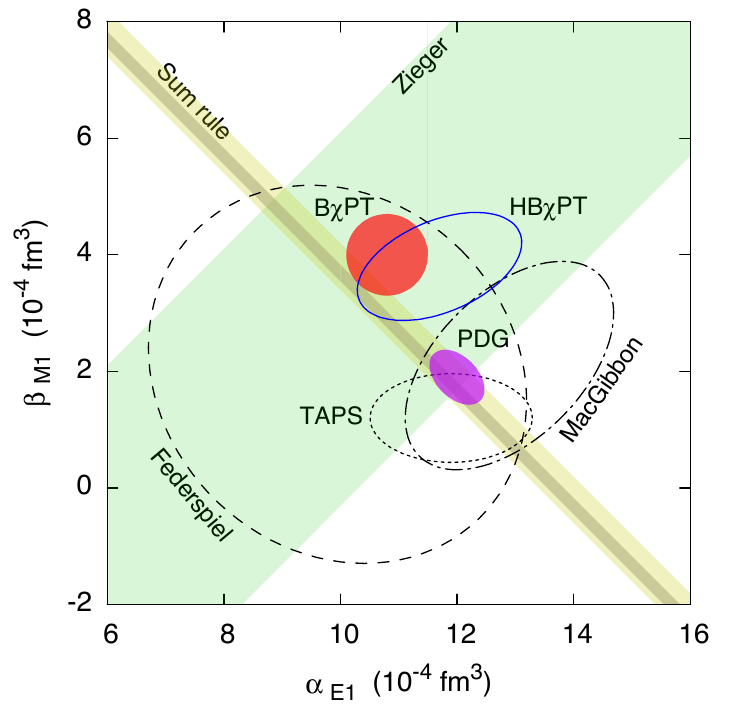}
\caption{(Color online). The scalar polarizabilities of the proton. 
Magenta blob represents the PDG summary~\cite{Beringer:1900zz}.
Experimental
results are from Federspiel et~al.~\cite{Federspiel:1991yd},
Zieger et al.~\cite{Zieger:1992jq}, MacGibbon et al.~\cite{MacG95},
and TAPS~\cite{MAMI01}.
`Sum Rule' indicates the Baldin sum rule evaluations of 
$\alpha_{E1}+\beta_{M1}$~\cite{MAMI01} (broader band) and \cite{Bab98}.
ChPT calculations are from \cite{Lensky:2009uv} (B$\chi$PT---red blob)
and the `unconstrained fit' of \cite{McGovern:2012ew} (HB$\chi$PT---blue ellipse).
} 
\figlab{potato}
%\end{minipage}
\end{figure}

A likely source of these discrepancies is an
underestimate of {\it model dependence} in the extraction
of polarizabilities from Compton-scattering data.
In principle, one should opt for a model-independent extraction, 
based on the low-energy expansion (LEX) of Compton-scattering
observables, where the leading-order terms, beyond the Born term, are expressed through
polarizabilities. For example,  the non-Born (NB) part
of the unpolarized differential cross section for
Compton scattering off a target with mass $M$ and charge $Ze$
is given by \cite{Baldin}
\bea
\frac{d\si^{\mathrm{(NB)}} }{d\Omega_L}  &=&-   \frac{Z^2\alpha_{em}}{M} 
\left(\frac{\nu'}{\nu} \right)^2 \nu \nu' \big[ \alpha_{E1} \left( 1 + \cos^2\th_L \right) \nn\\
&+& 2
\beta_{M1} \cos\th_L \big] + O(\nu^4),
\eea
where $\nu = (s-M^2)/2M$ and $\nu' =(-u+M^2)/2M$
are, respectively, the energies of the incident and scattered photon
in the lab frame, $\th_L$ ($d\Omega_L=2\pi \sin\th_L d\th_L$) is the scattering (solid) angle;
$s$, $u$, and $t=2M(\nu'-\nu) $ are the Mandelstam variables; and $\al_{em}=e^2/4\pi$ is the fine-structure constant.
Hence, given the exactly known Born contribution \cite{Powell:1949}
and the experimental angular distribution at very low energy,
one could in principle extract the polarizabilities with a negligible
model dependence. In reality, however, in order to resolve
the small polarizability effect in the tiny Compton cross sections,
most of the measurements are done at energies exceeding 100 MeV, 
i.e.,
not small compared to the pion mass $m_\pi$. It is $m_\pi$,
the onset of the pion-production branch cut, that
severely limits the applicability of a polynomial expansion in energy
such as LEX. At the energies around the pion-production
threshold one obtains a very substantial 
sensitivity to polarizabilities but needs to resort to a model-dependent
approach in order to extract them (see~\cite{Drechsel:2002ar,Schumacher:2005an} for reviews).

The magnetic polarizability $\be_{M1}$ seems to be affected the most:
the central value of the BChPT 
calculation is a factor of 2 larger than the PDG value. 
This is attributed  to the dominance
 of $\al_{E1}$ in the unpolarized cross section.\footnote{The problem
  is quite similar to the case of proton form factors, where the angular 
 (Rosenbluth)  separation from unpolarized scattering
 did not yield the correct result for the electric 
 form factor, due to the dominance of the magnetic contribution, and only
 separating the electric form factor from the magnetic one 
 by using polarization has yielded a break through. See \cite{HydeWright:2004gh,Perdrisat:2006hj} for reviews.} 
It is desirable to find an observable
sensitive to $\be_{M1}$ alone, such that
the latter could be determined independently
of $\al_{E1}$. 
According to the leading-order (LO) LEX for cross sections involving
linearly polarized photons~\cite{Maximon:1989zz}, the 
difference
of cross sections for photons polarized perpendicular
or parallel
to the scattering plane,
\beq
(d\si_{\perp} - d\si_{||})/d\Omega
\eqlab{alphaComb}
\eeq
 depends only on $\alpha_{E1}$, while
the  combination
\beq
( \cos^2\th \, d\si_\perp - d\si_{||} )/d\Omega
\eqlab{betaComb}
\eeq 
only on $\beta_{M1}$.
New experiments at the Mainz Microtron (MAMI) 
and the High Intensity Gamma Source (HIGS)   are planned
to measure these two combinations in order to extract 
$\alpha_{E1}$ and $\beta_{M1}$ independently.
This Letter aims to show that $\beta_{M1}$ can directly be extracted
from the beam asymmetry,
\beq
\Si_3 \equiv \frac{d\si_{||} - d\si_\perp}{d\si_{||} + d\si_\perp},
\eqlab{BSA}
\eeq
and that such extraction is potentially
more accurate than the one based on the observable given by \Eqref{betaComb}.

Indeed, applying the LEX for the beam asymmetry of proton
Compton scattering 
 we arrive at the following result:
\beq
\eqlab{leadingLEX}
 \Sigma_3 = \Sigma_3^{(\mathrm{B})} - 
\frac{4 M \omega^2 \cos\th \sin^2\th  }{\al_{em} (1+\cos^2\th)^2}\,\beta_{M1} 
+O(\w^4),
\eeq
where $\Sigma_3^{(\mathrm{B})} $ is the exact Born contribution, while
\bea
\omega &=& \frac{s-M^2 + \half t}{\sqrt{4 M^2 - t}}, \quad
\theta = \arccos\left(1 + \frac{t}{2 \omega^2} \right)
\eea 
are the photon energy and
scattering angle in the Breit (brick-wall) reference frame. 
In fact, to this order in the LEX the formula
is valid for $\w$ and $\th$ being the energy and angle in 
the lab or center-of-mass frame.

Equation \eref{leadingLEX} shows
that the leading (in LEX) effect of the electric polarizability
cancels out, while the magnetic polarizability
remains. Hence, our first claim is that 
a low-energy measurement of $\Si_3$ can in principle 
be used to extract $\beta_{M1}$ independently
of $\al_{E1}$, just as it was proposed for
 the combination of polarized cross sections given in
\Eqref{betaComb}. 
 
 In reality the low-energy Compton experiments on the proton are difficult because of small cross sections and overwhelming QED backgrounds.
 Precision measurement only becomes feasible for photon-beam energies
 above 60 MeV and scattering angles greater than 40 degrees.
 The upcoming experiments at HIGS and MAMI are planned
 for the energies between 80 and 150 MeV. As mentioned above, at these
 energies the effect of higher-order terms may become substantial.
 One way to see it is to compare the LEX result with the dispersion-relation calculations or calculations based on chiral perturbation theory.

 %\begin{widetext}

 \begin{figure}[b]
\includegraphics[width=\linewidth]{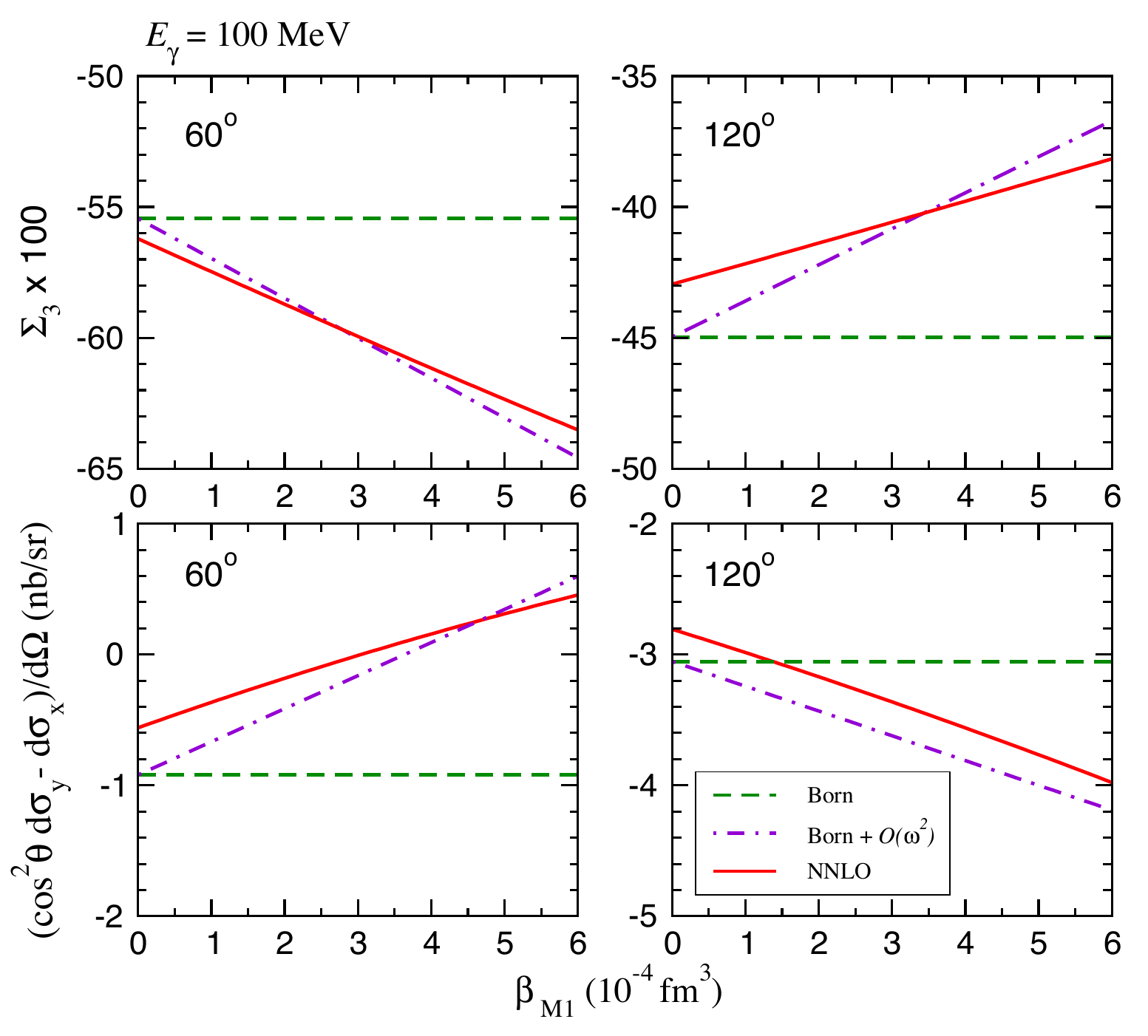}
\caption{(Color online). Beam asymmetry 
$\Si_3$ --- upper panel, and 
the linear combination of polarized cross sections
defined in \Eqref{betaComb} --- lower panel, shown
as function  of $\be_{M1}$ for fixed photon energy of 100 MeV
and scattering angles of 60 (left panels) and 120 (right panels) degrees.
The curves are as follows: dashed green --- Born contribution; dash-dotted magenta ---
the leading LEX formula \Eqref{leadingLEX}; red solid --- NNLO BChPT
\cite{Lensky:2009uv}.}
\figlab{E100}
%\end{minipage}
\end{figure}

\begin{figure}[t]
\includegraphics[width=\linewidth]{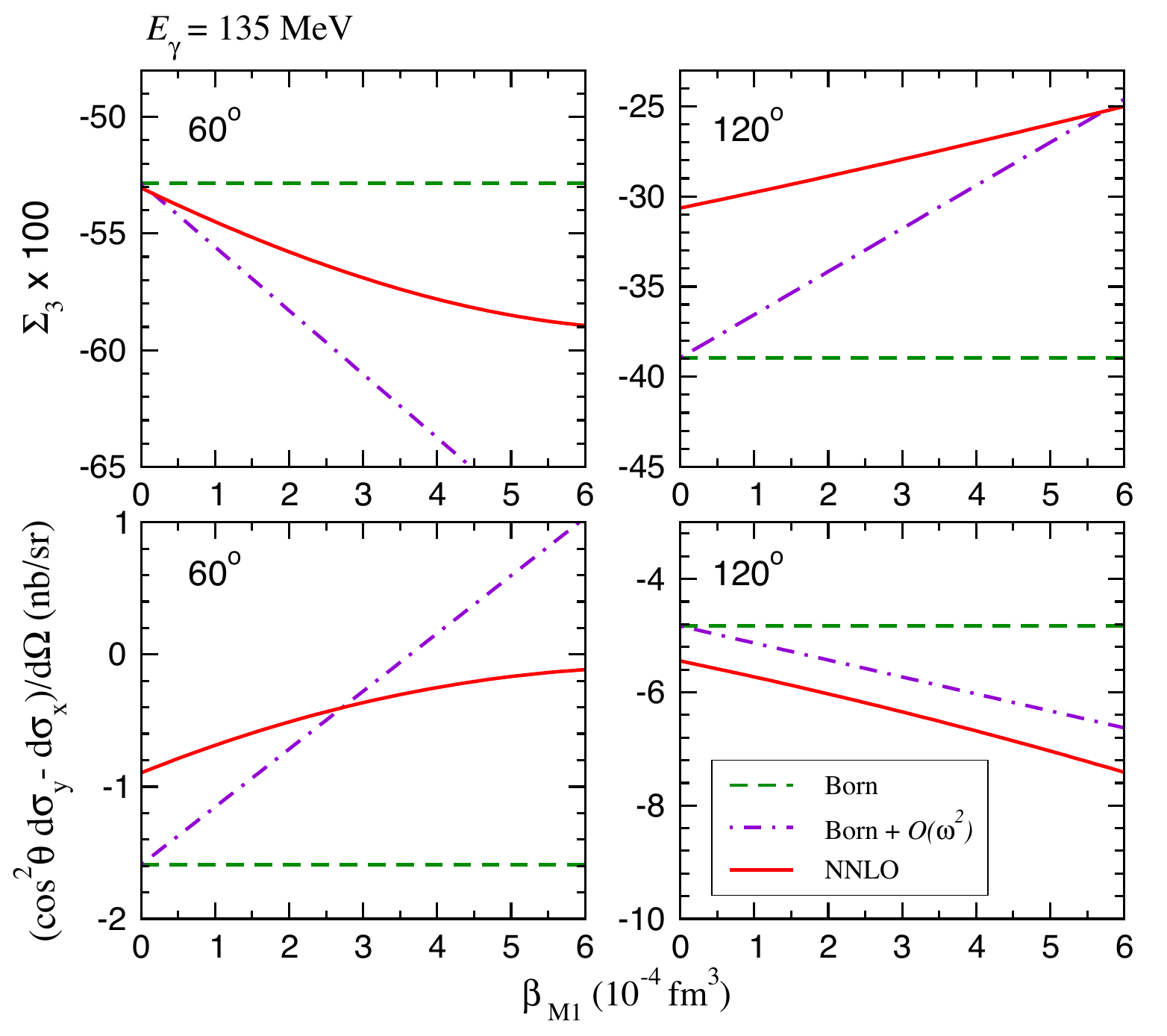}
\caption{(Color online). The same as in the previous figure but
for photon beam energy of 135 MeV.}
\figlab{E130}
%\end{minipage}
\end{figure}

 %\end{widetext}

 Figures~\ref{fig:E100} and \ref{fig:E130} demonstrate such a comparison of
 the leading-LEX result  to the next-next-to-leading order (NNLO) 
 BChPT result of Ref.~\cite{Lensky:2009uv} for the two observables
 defined in Eqs.~\eref{betaComb} and \eref{BSA}. The observables
 are plotted for the case of proton Compton scattering as a function of magnetic polarizability of the proton.
 From \Figref{E100} one sees that for the beam energy of 100 MeV
 the LEX is in a good agreement with the BChPT result, especially for the
 forward directions (left panels). 
 
 As expected we observe a significant
 sensitivity of these observables to $\be_{M1}$. However, the conventional 
 observable \Eqref{betaComb} is small compared to the unpolarised
 cross section (a few nano-barns per steradian compared to a few tens of nb/sr). It is therefore extremely difficult
 to measure it to the accuracy of few percent
 required here to discriminate between the PDG and ChPT values for
the magnetic polarizability displayed in \Figref{potato}. 
In contrast, the beam asymmetry is large and, given the fact that 
many systematic errors tend to cancel out in this observable, the required accuracy should be much easier to achieve. Still,  very high-intensity 
photon beams would be required to achieve the statistics necessary
to pin down the magnetic polarizability
model-independently to the accuracy currently claimed by the PDG, c.f.~\Eqref{PDGbeta}. The high-intensity electron facility MESA 
being constructed in Mainz is very promising in this respect.

 The results for the beam energy of 135 MeV (\Figref{E130})  show that
 the leading LEX result does not apply at such energies. 
 The differences
 between the LEX and BChPT curves at both 100 and 135 MeV 
 are mainly due to $\al_{E1}$ (which here is equal to
 zero for the LEX and about 11 for the BChPT) and the $\pi^0$-anomaly
 contribution.  While it is well known that the anomaly contribution is 
 proportional to $t/(t-m_\pi^2)$ and hence is maximal at the backward angles, 
 the $\al_{E1}$ contribution needs to be examined.

 In order to uniquely
define  the `recoil corrections' due to the polarizabilities,  we ought to
introduce them in a Lorentz-invariant fashion by
writing down an effective Lagrangian that yields the right
Hamiltonian in the static limit, i.e.,
\bea
\eqlab{effL}
\lag_{NN\ga\ga} &=& \pi \beta_{M1} \ol N  N  F^2\, \\
&-& \,
\frac{2\pi(\alpha_{E1}+\beta_{M1} )}{M^2} (\pa_\al \ol N)(\pa_\be N) F^{\al\mu} F^{\be\nu} \eta_{\mu\nu}\nn
\eea
where $F_{\mu\nu} = \pa_{[\mu} A_{\nu]}$ is the electromagnetic field-strength tensor,  $N(x)$ is the nucleon Dirac-spinor field, 
$\eta_{\mu\nu}
= \mathrm{diag} (1,-1,-1,-1)$ is the Minkowski metric.
Recalling that 
$
F^2= -2(F^{0i})^2 +  (F^{ij} )^2 =-2 \bE^2 + 2 \bB^2
$,
assuming the nucleon rest frame: $\pa_i N=0$,
$i \ga^0 \pa_0 N = M N $, $i \pa_0 \ol N \ga^0= -M \ol N $,
we obtain
\bea
\lag_{NN\ga\ga} &= & 2\pi \left\{
\beta_{M1} (\bB^2 -  \bE^2) 
+  (\alpha_{E1}+\beta_{M1})\,\bE^2 \right \}\ol N  N , \nn
\eea
which readily reproduces the well known 
nonrelativistic Hamiltonian: $4 \pi ( 
-\half \alpha_{E1} \bE^2 -\half \beta_{M1} \bB^2) $.

\begin{widetext} 
The Feynman amplitude corresponding to the Lagrangian
in \Eqref{effL} can be written as
\bea
{\cal M}_{\mathrm{NB}}^{\mu\nu} \veps_\mu^\prime \veps_\nu &=&
4\pi \, \ol u(p') \, u(p)  \big[    \beta_{M1} ( q\cdot q' \, \veps'\cdot \eps
- q\cdot \veps' \, q'\cdot \veps ) 
%\nn\\& +&
+\frac{\alpha_{E1}+\beta_{M1} }{2M^2}( p_\al' p_\be+ p_\al p_\be')\,  (q^{\prime\al} \veps^{\prime\mu}
- q^{\prime\mu} \veps^{\prime\al} ) (  q^\be \veps_{\mu}
- q_\mu \veps^{\be})
\big] \nn\\
&=& \ol u(p') \, u(p) \left[
- A_1^{(\mathrm{NB})}(s,t) \,  \cE' \cdot \cE + 
A_2^{(\mathrm{NB})}(s,t) \,  q\cdot \cE'\, q' \cdot \cE \right],
\eea
where $p$ and $q$ ($p'$ and $q'$) are the four-momenta of incident (outgoing) nucleon and photon and the manifestly gauge-invariant
polarization vectors are then
%\begin{subequations}
\beq
\cE_\mu  = \veps_\mu - \frac{(p'+p)\cdot \veps}{(p'+p)\cdot q} \, q_\mu\, ,
\quad\quad
\cE_\mu'  = \veps_\mu' - \frac{(p'+p)\cdot \veps'}{(p'+p)\cdot q} \, q_\mu'\, .
\eeq
%\end{subequations}
The polarizability contribution to the invariant 
Compton amplitudes is thus given as follows:
\begin{subequations}
\bea
A_1^{(\mathrm{NB})}(s,t) &=& 2\pi  (\alpha_{E1}+\beta_{M1} )(\nu^2 +\nu^{\prime 2}) 
+ 2\pi  \beta_{M1} t\, ,
\\
A_2^{(\mathrm{NB})}(s,t) &=& -4\pi \beta_{M1}\,.
\eea
\end{subequations}
We note that the contribution of $\alpha_{E1}+\beta_{M1}$  to $A_1$ differs from conventional
definitions by terms of higher order in the Mandelstam variable $t$, and, hence in energy. For instance, the difference of the present $A_1^{\mathrm{(NB)}}$ with the  
one in Ref.~\cite{Griesshammer:2012we} is equal to $- \pi (\alpha_{E1}+\beta_{M1} )\,  t(\w^2-\quarter t) $.
It turns out, however that this difference does not affect the
NLO contribution of polarizabilities to the beam asymmetry,
which together with the LO contribution of \Eqref{leadingLEX}
reads
%\begin{widetext}
\bea
\Sigma_3&=&\Sigma_3^{(\mathrm{B})} -
\,\frac{4 \cos\th \sin^2\th  }{ (1+\cos^2\th)^2}\, \frac{M^3 \beta_{M1}}{\al_{em}} 
\left(\frac{\omega}{M}\right)^2 \left\{ 1 + \left(\frac{\omega}{M}\right)^2\left[ \, a_1(\cos\th)
+\frac{M^3 \al_{E1} }{\alpha_{em}} \, \right]\right\}  \nn\\
&+& \frac{\sin^2\th}{  (1+ 
\cos^2\th)^2} \left(\frac{\w}{M}\right)^4 \left[ \, 
a_2(\cos\th)\, \frac{M^3\alpha_{E1}}{\alpha_{em}} 
+ a_3(\cos\th)\, \left(\frac{M^3 \be_{M1} }{\alpha_{em}} \right)^2\, \right]
+ O(\w^6),
 \eea
where the dimensionless coefficient functions are given by
(with $\kappa$ denoting the anomalous magnetic moment)
\bea
a_1(z) &=&    -\frac{1}{4(1+z^2)} \left[7-8 z+z^4 +2  \left(8 -16z +7 z^2 -z^4\right)\kappa  +
   \left(32 -40z -z^2 -z^4\right)  \kappa ^2 \right. \nn \\
  &+& \left. 4 
   \left(5-4z -3 z^2\right) \kappa ^3 + \left(5-3 z^2\right) \kappa ^4\right] ,\\
 a_2(z) & = &2 (1-z) +2  \left(2-4z+z^2\right) \kappa
+ \left(7-10 z - 2 z^2\right) \kappa ^2+4 
   \left(1-z-z^2\right) \kappa ^3 +  \left(1-z^2\right)\kappa ^4, \nn \\
    a_3(z) &=&  \frac{2(1-6 z^2+z^4)}{1+z^2}. \nn
\eea
 \end{widetext}

At order $\w^4$ there are also contributions from higher polarizabilities,
such as the following four spin polarizabilities~\cite{Drechsel:2002ar}: 
$\gamma _{E1E1}$, $\gamma _{E1M2}$,
$\gamma _{M1E2}$, $\gamma _{M1M1}$.  
Spin polarizabilities
are dominated by the $\pi^0$ anomaly and, hence, their contribution is only
appreciable in the backward angles. 
There is also a contribution 
from the fourth-order scalar polarizabilities $\al_{E2}$, $\be_{M2}$, and
$\be_{M\nu}$, 
which can be obtained
by the following replacement  in the $O(\w^2)$ term:
\beq
\be_{M1}\to \be_{M1} + \w^2 \,\left( \be_{M\nu}
- \mbox{$\frac{1}{12}$} \al_{E2} + \mbox{$\frac{1}{6}$}
\be_{M2}\cos\th\right).
\eeq

Examining the $O(\w^4)$ contribution of scalar polarizabilities shown above,
we note that the coefficient functions containing $\kappa$ (i.e., $a_1$ and $a_2$) can become large in the backward angles. To limit the impact of
these terms, as well of the spin polarizabilities, on the extraction of $\be_{M1}$, 
the scattering angle should be chosen in the vicinity of 60 degrees. As is illustrated in \Figref{E100}, in this case the leading LEX result is very
close to the BChPT result, thus confirming the near-perfect cancellation
of higher-order terms in this case.

To conclude, the beam asymmetry
$\Si_3$ is instrumental in isolating the contribution
of the magnetic polarizability $\be_{M1}$ to low-energy Compton scattering. 
While the cross sections  receive contributions from both the
electric  and magnetic  polarizability, the effect of $\al_{E1}$
cancels out from the asymmetry at leading order in the low-energy expansion.
We have shown that the next-to-leading corrections are suppressed at the
forward scattering angles.  A precise 
and model-independent determination of
the proton $\be_{M1}$ is feasible through
a precision measurement of $\Si_3$ at beam energies
below 100 MeV and forward scattering angles.
Furthermore,  when multiplied with the unpolarized cross section, $\Si_3$
yields the polarized cross section difference, which
provides an exclusive access to the electric polarizability.
A measurement of the beam asymmetry at these low energies has
not been done yet, but is being planned at the MAMI and HIGS
facilities and could be done in the very near future.

\medskip
%\section*{Acknowledgements}

We thank Dave Hornidge and Judith McGovern 
for helpful remarks on the manuscript.
We are happy to acknowledge the support of the Deutsche Forschungsgemeinschaft (DFG) through Collaborative 
Research Center SFB 1044 ``The Low-Energy Frontier of the Standard Model"
and the Graduate School DFG/GRK 1581
``Symmetry Breaking in Fundamental Interactions".

%\end{document}

\end{document}